\documentclass[preprint]{aastex6}
\begin{document}

\title{CALSPEC: WFC3 IR GRISM SPECTROPHOTOMETRY}

\author{Ralph~C.\ Bohlin\altaffilmark{1}, and Susana E. Deustua\altaffilmark{1}}
\altaffiltext{1}{Space Telescope Science Institute, 3700 San Martin Drive,
Baltimore,  MD 21218, USA}

\begin{abstract}

The collections of spectral energy distributions (SEDs) in the \emph{Hubble
Space Telescope} (HST) CALSPEC database are augmented by 19 IR SEDs from Wide
Field Camera 3 (WFC3) IR grism spectra. Together, the two IR grisms, G102 and
G141, cover the 0.8--1.7~\micron\ range with resolutions R=200 and 150,
respectively. These new WFC3 SEDs overlap existing CALSPEC Space Telescope
Imaging Spectrograph (STIS) standard star flux distributions at 0.8-1~\micron\
with agreement to $\lesssim$1\%. Some CALSPEC standards already have near-IR camera and multi-object spectrogragh (NICMOS) 
SEDs; but in their overlap region at  0.8--1.7~\micron, the WFC3 data have
better wavelength accuracy, better spectral resolution, better repeatability,
and, consequently, better flux distributions of $\sim$1\% accuracy in our
CALSPEC absolute flux SEDs vs. $\sim$2\% for NICMOS. \textbf{With the improved
SEDs in the WFC3 range, the modeled extrapolations to 32~\micron\ for JWST flux
standards begin to lose precision longward of the 1.7~\micron WFC3 limit,
instead of at the 1.0~\micron\ long wavelength limit for STIS. For example, the
extrapolated IR flux longward of 1.7~\micron\ for 1808347 increases by $\sim$1\%
for the model fit to the data with WFC3, instead of just to the STIS SED alone.}
\end{abstract}


\section{Introduction}			%S1

Absolute flux distributions of standard stars with pedigrees traceable to the
International System of Units (SI) are essential for 21st century astrophysics
and precision cosmology. For example, dark energy investigations that rely on
observations of Type Ia supernovae (SNe Ia) require an absolute color
calibration, i.e. band-to-band, with an accuracy of better than 1\%
\citep{scolnic14,stubbs15}. JWST, WFIRST and other future large space
telescopes, along with large ground-based projects like PanSTARRS and LSST, need
reliable flux standard stars with a wide range of brightness over the whole sky
\citep{stubbs16}. 

Stellar-flux uncertainty requirements are driven in part by the use of SNe Ia as
probes of the expansion history of the universe. In 1998, the accelerating
expansion of the universe was discovered, which implies the existence of a new
component of the universe called dark energy. Cosmological and dark energy
parameters are determined from the SNe Ia brightness-redshift relationship
(Hubble diagram). To distinguish between competing dark energy models whose
predictions differ by as little as 2\% \citep{albrecht2006}, dark energy
investigations require sub-percent absolute spectral radiometric uncertainty and
cross-calibration of the relative zero-points of all bands to an accuracy of
better than 0.5\% over the wavelength range of 0.48--2\micron.
Therefore,missions like WFIRST will depend upon a network of primary standard
stars measured with absolute photometric uncertainty of better than 1\%
(0.01mag) over this range, which is an ambitious but achievable goal.

Although the motivation for sub-percent flux calibration accuracy is driven by
cosmology, reducing the uncertainties in absolute stellar spectrophotometry
benefits other research areas. The fundamental parameters of stars, including
mass, radius, metallicity, and age, are inferred by matching accurate models of
stellar atmospheres to calibrated spectroscopic data in order to determine the
effective temperature, surface gravity, composition, and interstellar reddening.
Model atmospheres for stars with relatively simple atmospheres, such as pure
hydrogen white dwarfs, combine with stellar interior models to predict
photometric parameters, stellar radii, and absolute luminosities. These results
plus precise photometry predict stellar distances, which are in excellent
agreement with measured trigonometric parallaxes \citep{Holberg2008}. When
combined with GAIA's exquisite astrometric accuracy, which yields parallaxes
with uncertainties of 0.04-0.7 milliarcseconds for 1.5 billion stars,
sub-percent spectrophotometry makes possible meaningful tests of 3-D spherical
stellar models. Masses are directly calculated, and quantitative tests of
evolutionary models are improved.

Currently, the best choice of fundamental standards in the UV to near-IR are the
CALSPEC models for the primary pure hydrogen white dwarfs (WDs) G191B2B, GD153,
and GD71, which are the basis for the spectral energy distributions (SEDs) in
the HST/CALSPEC\footnote{http://www.stsci.edu/hst/observatory/crds/calspec.html}
database. This paper specifies procedures required to place WFC3 IR grism SEDs
on the CALSPEC absolute flux scale of \citet{bohlinetal14}. Section 2 outlines
the data processing steps required to extract photometric spectra from the  WFC3
IR grism images. Section 3 details the correction for the changing sensitivity
with time, while Section 4 compares WFC3 SEDs with the STIS results from CALSPEC
in order to derive the WFC3 count rate non-linearity (CRNL) correction. Section
5 summarizes the absolute flux calibration of the WFC3 IR grisms; and in Section
6, the fully corrected WFC3 SEDs are compared with models to extend predictions
of the flux distributions to 32 \micron\ for JWST calibration purposes.

\section{Data Analysis} %S2

\subsection{The Data}	%2.1

\textbf{The publically available WFC3 IR grism observations of CALSPEC stars in
the nominal Stare Mode include the white dwarf standard stars G191B2B, GD153,
and GD71, plus sixteen other CALSPEC stars. These spectra were acquired at
multiple locations on the IR detector. Grism Scanned Mode observations of
CALSPEC stars will be discussed in a subsequent publication (Deustua \& Bohlin,
in prep). Table~\ref{table:obs} itemizes the 19 WFC3 grism observations in Stare
Mode, and indicates whether NICMOS observations are also available.}

\begin{deluxetable}{lrrlccc}	    %Table1
\tablewidth{0pt}
\tablecolumns{6}
\tablecaption{\label{table:obs} WFC3 IR Observations of CALSPEC Stars}
\tablehead{
\colhead{Star} &\colhead{CALSPEC Name} &{J\tablenotemark{a}}  &{Sp.Ty.\tablenotemark{a}} 
&{G102\tablenotemark{b}} &{G141\tablenotemark{b}} &\colhead{NICMOS}}
\startdata
2MASS J17571324+6703409 &1757132       &11.31  &A3V  &8  &8   &No   \\
2MASS J18022716+6043356 &1802271       &11.87  &A2V  &6  &8   &Yes  \\
2MASS J18083474+6927286	&1808347       &11.65  &A3V  &6  &3   &No   \\
2MASS J00361617+1821104 &2m003618      &12.47  &L3.5  &2  &2   &Yes  \\
2MASS J05591914-1404488 &2m055914      &13.80  &T4.5  &2  &233 &Yes  \\
BD+60 1753              &bd60d1753     &9.61   &A1V  &8  &8   &No   \\
2MASS J03323287-2751483 &c26202        &15.40  &F8IV  &0  &4   &Yes  \\
G191B2B                 &g191b2b       &12.54  &DA.8  &19  &15 &Yes  \\
GD153                   &gd153	       &14.01  &DA1.2  &77  &77 &Yes  \\
GD71                    &gd71	       &13.73  &DA1.5  &58  &58 &Yes  \\
GRW+70$^{\circ}$5824    &grw\_70d5824  &13.25  &DA2.4  &26  &27 &Yes  \\
HD37725                 &hd37725       &7.95   &A3V  &1  &0   &No   \\
2MASS J17583798+6646522	&kf06t2        &11.90  &K1.5III  &3  &4   &Yes  \\
2MASS J16313382+3008465	&p330e	       &11.77  &G2V  &8  &13  &Yes  \\
2MASS J16194609+5534178 &snap2	       &14.97  &G0-5  &1  &2   &Yes  \\
VB8                     &vb8	       &9.78   &M7V &2  &2   &Yes  \\
WD1327-083              &wd1327\_083   &12.62  &DA3.5  &2  &2   &No   \\
WD1657+343              &wd1657\_343   &...    &DA.9  &1  &1   &Yes  \\
WD2341+322              &wd2341\_322   &13.17  &DA3.8  &2  &2   &No   \\
\enddata
\tablenotetext{a}{J mag and Spectral Type are from Simbad}
\tablenotetext{b}{Number of available WFC3 observations that are included
in our average SEDs.}
\end{deluxetable}

\subsection{Extraction of Spectra from the Images}	%2.2

The software for producing files of flux vs.
wavelength for the WFC3 IR grisms is an adaptation of similar code that was
written by D. J. Lindler and utilized for the NICMOS grism data
\citep{bohlin05nic,bohlin06}. Our procedure takes advantage of the zero order method \citep{bohlin2015}.
The first
step is to locate the zero-order image and establish the wavelength scale.
This method requires an accuracy of 
$\sim$1\arcsec\ for the astrometry in the data file headers. In case the
observer specified incorrect coordinates or omitted a significant proper motion,
the target coordinates are updated in the input file headers. 

If there is no zero order on the grism image, but there is an associated direct
image, the (x,y) position of the star is found in the direct image, Then, the
AXE method \citep{kuntschner2009a, kuntschner2009b} is used with updated
dispersion constants \citep{pirzkal2016} and wavelengths referenced to the
stellar position on the direct image. In the rare case where neither a direct
image nor a zero order is available, then the astrometry is often good enough to
locate the spectrum, e.g. icqw01b1q for GD71, where a large wavelength
correction of 91~\AA\ is required.

To establish the spectral trace to better precision than predicted by the
astrometry, the image is searched for the -1, +1, and +2 orders. A linear least
square fit to the (x,y) positions of the zero order and to any of the three
spectral orders that are found determines the exact slope and location of the
spectral trace. A discussion of the range of measured slopes appears in
\citet{bohlin2015}. 

After establishing the wavelength vector and spectral trace of the spectrum's
position on the grism image, a scaled master sky background is subtracted to
remove the structure in the vignetted sky background. Next, a
wavelength-dependent flat-field data-cube (Deustua 2019, in prep) is applied to
the relevant pixels in the image.

The final processing steps extract the gross spectrum with a height of six
pixels ($\sim$0.78\arcsec) and subtract the residual background to produce the
net stellar signal vector. The six pixel height is a trade-off between
collecting more signal for bright stars and minimizing excess noise for faint
sources. This choice includes 85-90\% of the total signal
\citep{kuntschner2009a, kuntschner2009b}. To maintain photometric
accuracy with the relatively small height of six pixels, the spectral trace must
be located with a precision of $\lesssim$1 pixel. Section 5 covers the absolute
flux calibration.

\begin{figure}			%fig1
\centering 
\includegraphics*[width=.9\textwidth,trim=40 0 0 0]{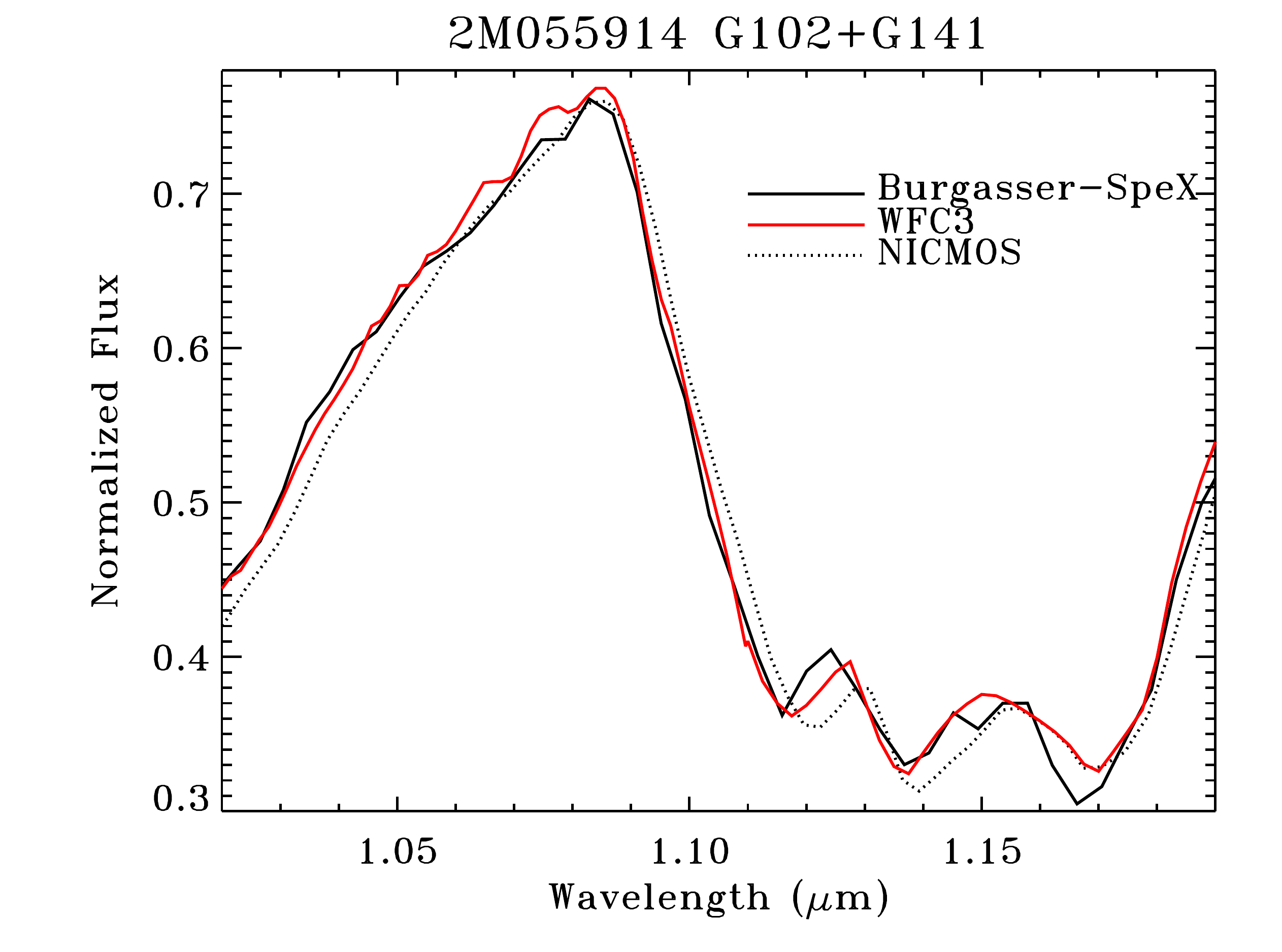}
\caption{\baselineskip=12pt
A section of the spectrum of 2M055914 from WFC3 (red) overplotted with the
NICMOS spectrum (dotted) and SpeX data of \citep{burgasser06} (solid black).
The WFC3 agrees much better with SpeX, while NICMOS is shifted significantly
toward longer wavelengths.
\label{spex}} \end{figure}

\textbf{The WFC3 IR grisms have dispersions of 24.5 \AA\ per pixel for G102 and
46.5 \AA\ per pixel for G141, and \citet{bohlin2015} quote a wavelength
precision of 5 and 9~\AA\ for G102 and G141, respectively. To verify the
wavelength accuracy, a comparison of the  SpeX spectrum of the brown dwarf star
2MASS J05591914-140448 (2M055914 in the CALSPEC database) with WFC3 IR is shown
in Figure~\ref{spex}. The SpeX spectrum of \citep{burgasser06} has a spectral
resolution of $\sim$150 between 0.8 and 2.5 micron, comparable to WFC3 IR, and a
wavelength precision of $\pm0.3$~\AA\ \citep{Cushing2004}. {Figure~\ref{spex}}
demonstrates that the WFC3/IR wavelength scale agrees with SpeX to within
10~\AA, while the NICMOS spectrum of the same source shows differences of up to
40~\AA. NICMOS used three grisms to cover the wavelength range between 0.8 to
2.5 \micron\ with dispersions of 50-100 \AA\ per pixel depending on the
grism\footnote{At present, NICMOS is not an active HST instrument. See
http://www.stsci.edu/hst/nicmos.}.}

\subsection{Adjustment of the Wavelength Scales}	%2.3

\textbf{In order to obtain accurate flux calibrations, a precise wavelength
scale must be established; but the extraction of a spectrum from its image is
subject to small wavelength errors due to inaccuracies in either finding the
zero order in the grism image or the stellar image on the reference direct
image. When the zero order is off the grism image and no direct reference image
is available, larger wavelength errors are expected. Therefore, the initial
results are updated by comparison to spectral features in the overlap region
with STIS between 0.8 and 1.0 $\micron$ (for G102) or with a reference spectrum
at longer wavelengths (for G141). For an A-star type spectrum, the strong
Paschen and Brackett hydrogen series provide clean isolated lines with precisely
known wavelengths. }

\subsection{Co-adding and Merging Individual Observations}	%2.3

All the separate G102 and G141 observations of a source are co-added and merged.
The default merge point where the co-add switches from G102 to G141 is at
11250~\AA. The fluxes of the two independent measurements agree within the
accuracy goal of 1\% at the merge point.

\section{The Correction for Changing Sensitivity} %S3

\begin{figure}			%fig2
\centering 
\includegraphics[width=5.5in,trim=0 50 0 0]{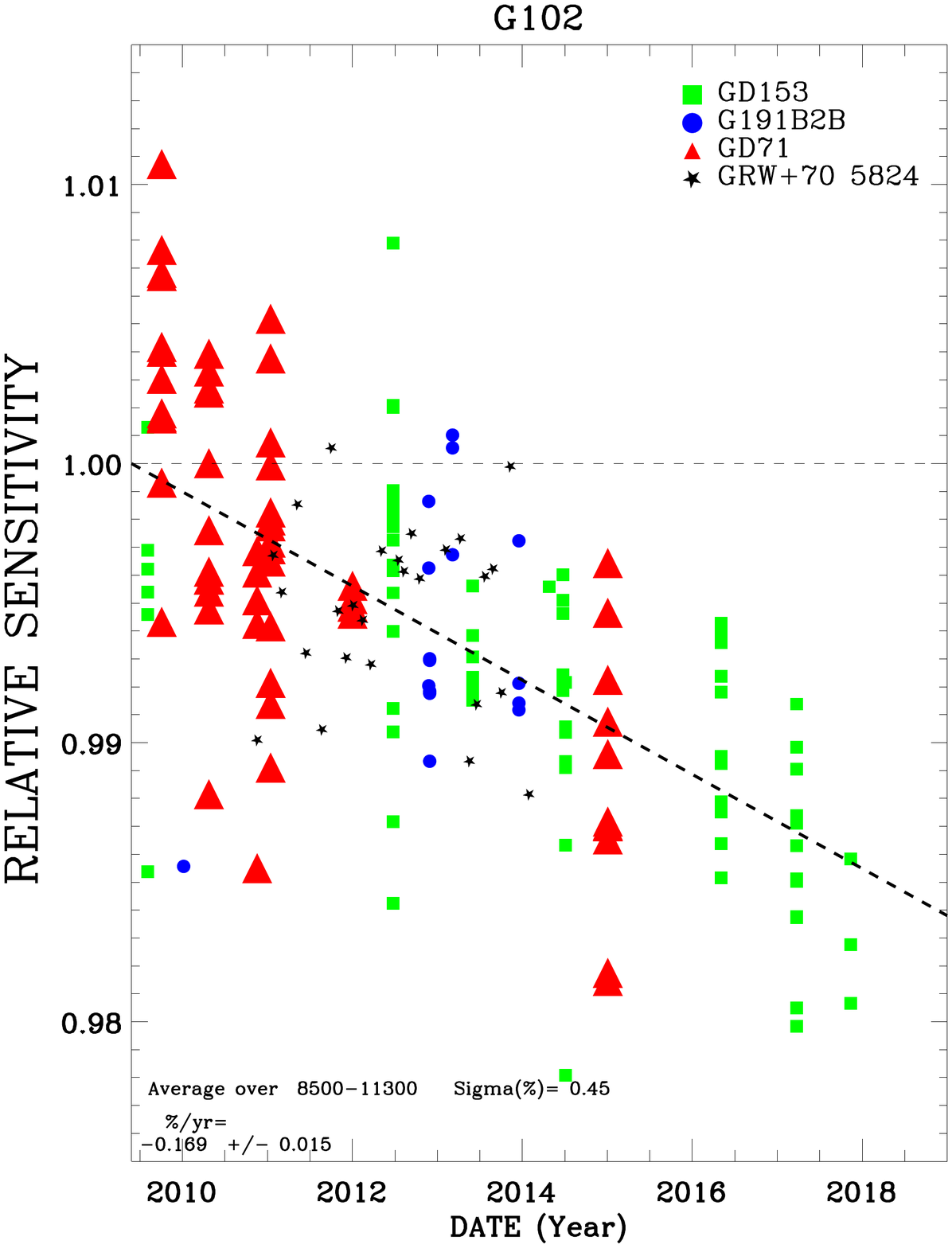} 
% \vspace*{-1.0 cm}
\caption{The change in sensitivity with time since 2009 for the WFC3 G102 grism after correction for non-linearity. The loss rate of $0.169 \pm 0.015\%~yr^{-1}$ and the rms redidual scatter of 0.45\% are written near the bottom of the plot.
The rms scatter is a measure of the repeatability, i.e. the 1$\sigma$ broadband uncertainty in an individual observation. The least sguare fit (dashed line) and the data points are all normalized to unity at 2009.4 by dividing by initial value of the fit at 2009.4.}
\label{var102} 
\end{figure}

\begin{figure}			%fig3
\centering 
\includegraphics*[width=5.5in,trim=0 50 0 0]{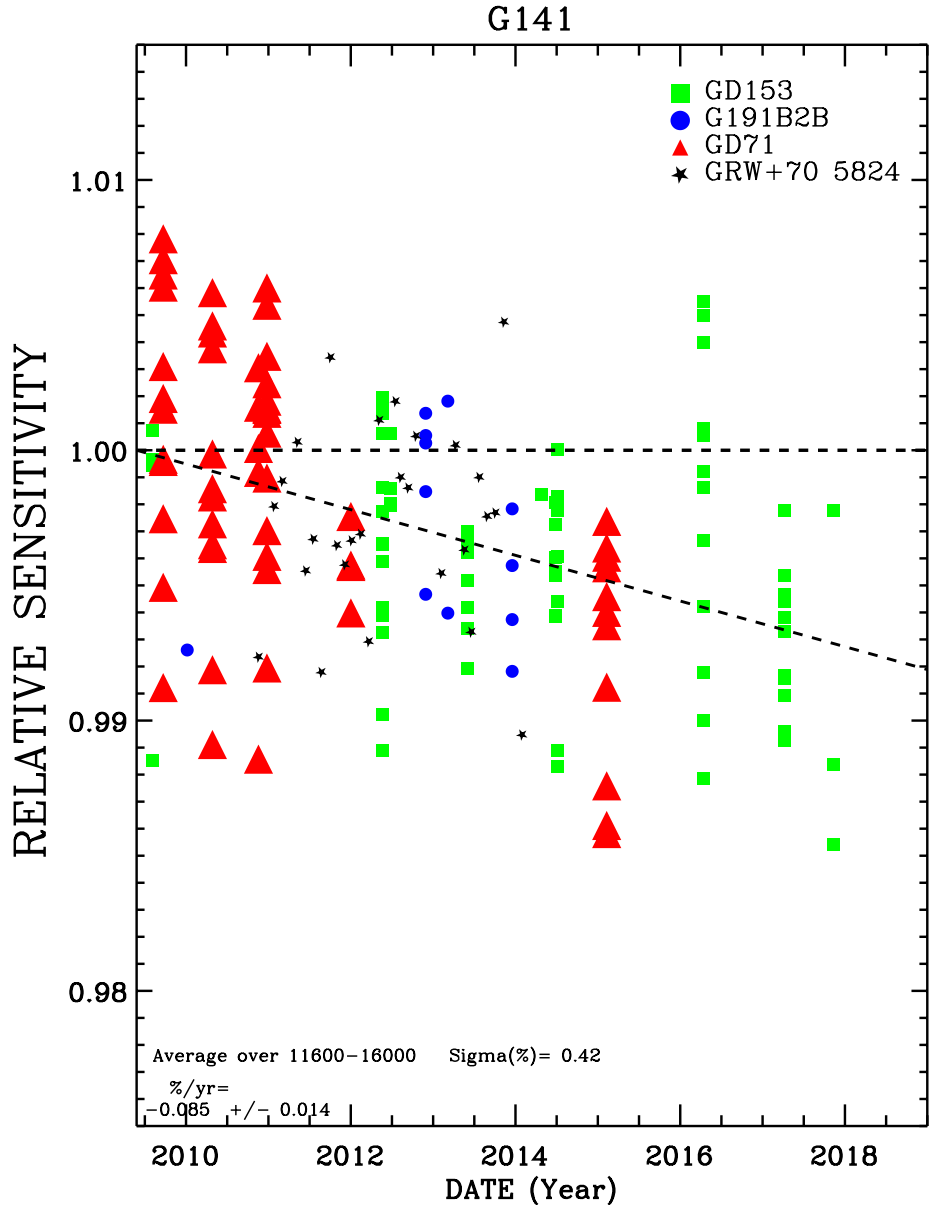}
\caption{The change in sensitivity with time for the WFC3 G141 grism after correction for non-linearity. The loss rate of $0.085 \pm 0.014\%~yr^{-1}$ and the rms scatter of 0.42\% are written near the bottom of the plot. The rms scatter is a measure of the repeatability, i.e. the 1$\sigma$ broadband uncertainty in an individual observation. The least sguare fit (dashed line) and the data points are all normalized to unity at 2009.4.}
\label{var141}
\end{figure}

\textbf{To investigate temporal sensitivity change, the spectral data of the
four monitor stars, G191B2B, GD153, GD71, and GRW+70$^{\circ}$5824, are binned
over the regions of peak sensitivity from 8500--11300~\AA\ for G102 and
11600--16000~\AA\ for G141. For each monitor star, the individual bins are
divided by their ensemble average. Figures~\ref{var102} and \ref{var141} show
all 177 ratios for G102 and all 174 ratios for G141 as a function of time.
Linear least-square fits to the ratios give slopes of $0.169 \pm
% 2019mar18 - updated post re-submission from 84 to 85:
0.015\%$ per year for G102 and $0.085 \pm 0.014\%$ per year for G141. The points
in Figures~\ref{var102} and \ref{var141} have small corrections for the
non-linearity from the next section. These sensitivity changes are likely due to
the polymerization of contaminants on the optical surfaces, which causes similar
results for ACS \citep{bohlinetal11}, where the loss rates for the long
wavelength filters F606W, F775W, and F814W are 0.26, 0.13, and
0.09\%, per year respectively. For STIS, the low dispersion grating mode G750L
shows losses in the range of  0.11--0.23\% per year for the 5500--8300~\AA\
wavelength range (Bohlin, unpublished). }

These standard star observations are spaced over the whole detector, which means
that the rms scatter includes uncertainties in the flat-field correction. The
precision of stellar SEDs at the same position is slightly better. For example,
the GRW+70$^{\circ}$5824 observations are all within 3.5\arcsec\ of center; and
the rms scatter is 0.35 and 0.37\% for G102 and G141, respectively, compared to
0.45\% and 0.42\% for the whole ensemble.

To check for a wavelength dependence of the sensitivity changes, a fit to the
region 10000--11600~\AA\  for G102 results in a slope of $0.192 \pm
0.019\%$ per year, which agrees within $\sim$1$\sigma$ of the adopted
$0.169\%$ per year. These time-dependent sensitivity corrections are applied to
the individual grism spectra extracted as described in Section 2.

As of 2018 December, the most recent set of WFC3 IR grism observations of the
four monitor stars was 2017 Nov 11. Routine monitoring is needed for the
valuable WFC3 grism modes. All four stars should be observed
in both IR grism modes every year, which is the cadence adopted by ACS for the
most important filters; STIS sets the cadence at every two years. A standard
reference position for the monitor stars improves the precision for
absolute flux measurements on the IR detector.

%\newpage

\section{Count Rate Non-linearity Correction for the WFC3 IR Grism Data} %S4

The WFC3 IR detector is a HgCdTe type, which is inherently non-linear. One type
of non-linearity is the count-rate non-linearity correction (CRNL) found for
NICMOS \citep{bohlin05nic, bohlin06}. CRNL should not be confused with the
total count non-linearity, which occurs as the charge acccumulated in a pixel
approaches the  detector full-well depth. The WFC3 HgCdTe detector is an
improved version but still must be investigated for a corresponding
non-linearity. In the case of NICMOS, the best measures of the non-linearity
were from comparisons to STIS below 1~\micron, to WD model SEDs above 1~\micron,
and to results from high  background observations of P014C. However, WFC3 cannot
superpose lamp illumination on a stellar observation  to raise the background,
which limits the present analysis to comparisons of the grism data to STIS and
to stellar atmoshpere models that match the STIS SEDs.

In addition to the high-background result, the NICMOS CRNL measures relied
heavily on models for the pure-hydrogen WDs WD1057+719 and WD1657+343 longward
of 1~\micron. However, WD1057+719 is not observed, and only one observation of
WD1657+343 exits in the present WFC3 IR grism dataset. The set of WFC3 IR SEDs
could be compared to the overlapping set of NICMOS results, but the uncertainty
in the NICMOS correction is greater than the total WFC3 CRNL. Instead,
WFC3 data are corrected for sensitivity changes with time and then compared
to stellar models that extend STIS observations into the IR using fits from the BOSZ
stellar atmosphere grid \citep{bohlin2017}. The models are fit to only the STIS
data without using NICMOS or WFC3 as a constraint in the IR. Then, both the WFC3
and model SEDs are binned in 500~\AA\ bins, and the flux ratios of WFC3 IR grism
to model are plotted against the WFC3 net count rate in electrons s$^{-1}$ for
each bin.

\begin{figure}[htb!]			%fig4 fullin.pro
\centering 
\includegraphics[width=4.75in]{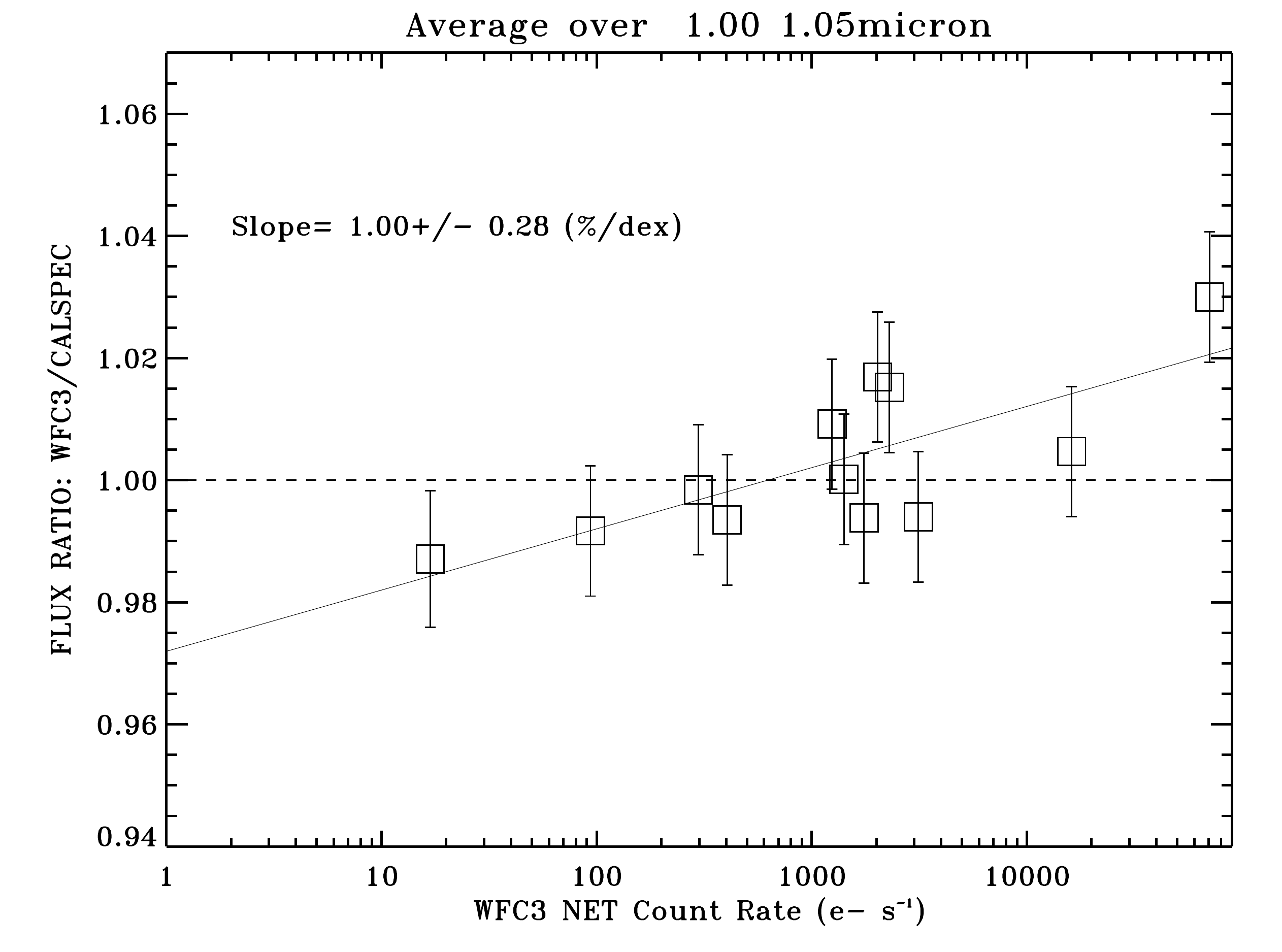}
\caption{The WFC3 G102 CRNL for the 1--1.05~\micron\ wavelength bin from a
comparison of the average grism SED with a model for each of the 12 stars. The
models are independent of any NICMOS or WFC3 data. The line is the least square
fit to the 12 data points and has a slope of $1.00\pm 0.28\%$ per dex. The
uncertainties shown as error bars are dominated by the repeatability of the 
grism observations in a 500\AA\ bin.}  \label{onebin} \end{figure}

An example of such a plot appears in Figure~\ref{onebin} for the
1--1.05~\micron\ bin. The slope of the least-square fit,
$1.00~\pm~0.28~\%$ per dex, is the measure of the CRNL, where the
faint stars are too faint and the bright stars too bright by 1.00\% for each
factor of 10 in the count rate. While there is a moderate amount of scatter
among the data points, the result for this one bin has a 4$\sigma$ level of
significance.

\begin{figure}[htb!]			%fig5
\centering 
\includegraphics[width=4.75in,trim=0 60 0 0]{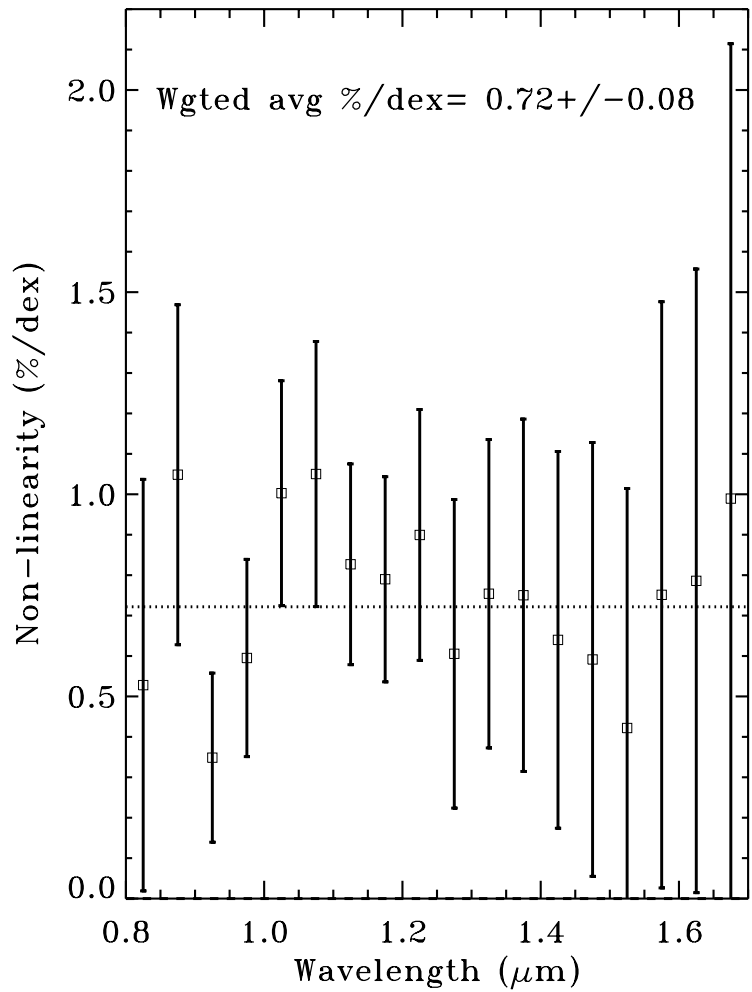} 
\caption{The WFC3 IR grism count rate non-linearity as a function of wavelength
with 1$\sigma$ error bars. None of the results among the 18 bins of 500~\AA\ (0.05~\micron) width differs from the average of $0.72~\pm 0.08~\%$ per dex by as much as 2$\sigma$, and only one bin differs by more than 1$\sigma$. The data points are
derived from comparisons of grism SEDs to models that are fit to STIS only, as in Figure~\ref{onebin}.}
\label{allbin}
\end{figure}

\textbf{Figure~\ref{allbin} shows the CRNL} for all 18 of the 500~\AA\
bins as a function of the mean wavelength of each bin. The error bars increase
with wavelength, because the modeled extrapolations of the SEDs lose precision
with increasing distance from the STIS anchor spectra below 1~\micron. Except
for the one bin centered at 9250~\AA, the count-rate non-linearity measures are
all within 1$\sigma$  of the weighted mean value of
$0.72\pm 0.08\%$ per dex.  

\textbf{A recent analysis using IR photometery finds $0.77\pm 0.08~mag$ per dex
\citep{riess2019}. Our result of $0.72\pm 0.08\%$ per dex from grism
spectrophotometry corresponds to $0.78~mag$ per dex. As an unfortunate result of
verbal communication, \citet{riess2019} understood our result to be in in mag
units rather than percent. The independent \citet{riess2019} result of
$0.77~mag$ per dex from all photometry corresponds to $0.71~\%$ per dex, and the
average of the two independent analyses should be $0.715~\%$ per dex or
$0.775~mag$ per dex . For this work, $0.72\pm 0.06\%$ per dex is adopted for
the average result.}

Statistically, this average value of the CRNL has a high level of
significance and is implemented in the WFC3 IR grism data processing as a
correction from the measured net count rate N(obs) to corrected N(corr) as
\begin{equation}N(corr)={N(obs) \over 1.0072 ^{log(N(obs)/1000)}}~~, \label{eql}
\end{equation} where the observed count rate is divided by 1000 to make the
correction near unity in the brightness range of the three primary WD reference
SEDs.

\section{The Absolute Flux Calibration} %S5

Because the spectra are uniformly extracted with a height of six pixels, 
sensitivities, i.e. flux calibrations, are required  only for this case.
Extracted signals in electrons~s$^{-1}$ are compared with the SEDs of the
primary standards G191B2B, GD153, and GD71 \citep{bohlinetal14} to define the
flux calibration \textit{S} with units of electrons~s$^{-1}$ per [erg~s$^{-1}$
cm$^{-2}$ \AA$^{-1}$]. The extracted signal is divided by \textit{S} to get the
absolute flux after correcting for the change in sensitivity with time from
Section 3 and for the CRNL from Section 4.

\begin{figure}[htb!]			%fig6
\centering 
\includegraphics*[width=0.61\textwidth,trim=0 50 0 0]{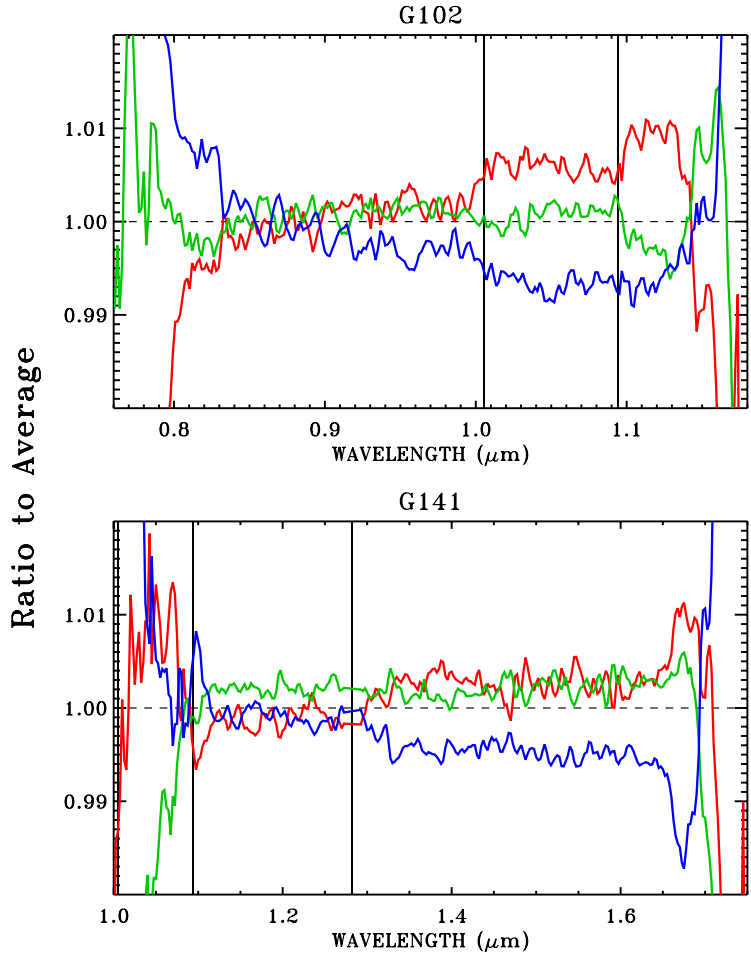}
\caption{Ratio of the sensitivities for each of the primary standards to
their average \textit{S} for the IR grisms G102 and G141. Red-G191B2B,
Green-GD153, Blue-GD71. Vertical lines mark the positions of the stronger absorption lines of
hydrogen P$\delta$, P$\gamma$, and P$\beta$, but notice the lack of spurious
features in the sensitivities at those wavelengths.}  \label{sens} \end{figure}

Each of the three primary standards has equal weight, and the ratio for each
star to final average \textit{S} is shown in Figure~\ref{sens}. The
sensitivities at the strong P$\beta$ line at 12822~\AA\ (vac) are interpolated
across. The calibrations for the separate stars agree with their average to
$<$1\% from 8100--11500~\AA\ for G102 and from 10700--16600~\AA\ for G141, which
demonstrates consistency to $<$1\% in our merged fluxes from 8100--16600~\AA.
The deviations of the curves from unity reflect the errors in the standard star
SEDs combined with flat field errors that may cause inconsistent average signals
due to the heterogeneous distributions of observation over the detector. Scatter
near the ends of the wavelength ranges is caused by small wavelength errors and
low signal that approaches the background level.

Table~\ref{table:sens} collates the sensitivies \textit{S} for G102 and G141,
while Table~\ref{table:wfcstub} contains the resulting collection of new WFC3 IR
grism SEDs.

\section{Extending the SEDs into the Mid-IR for JWST Calibration} %S6

\subsection{Comparison of WFC3 to STIS and NICMOS SEDs}		%6.1

\begin{figure}[htb!]			%fig7
\centering 
\includegraphics*[width=0.7\textwidth,trim=0 0 0 0]{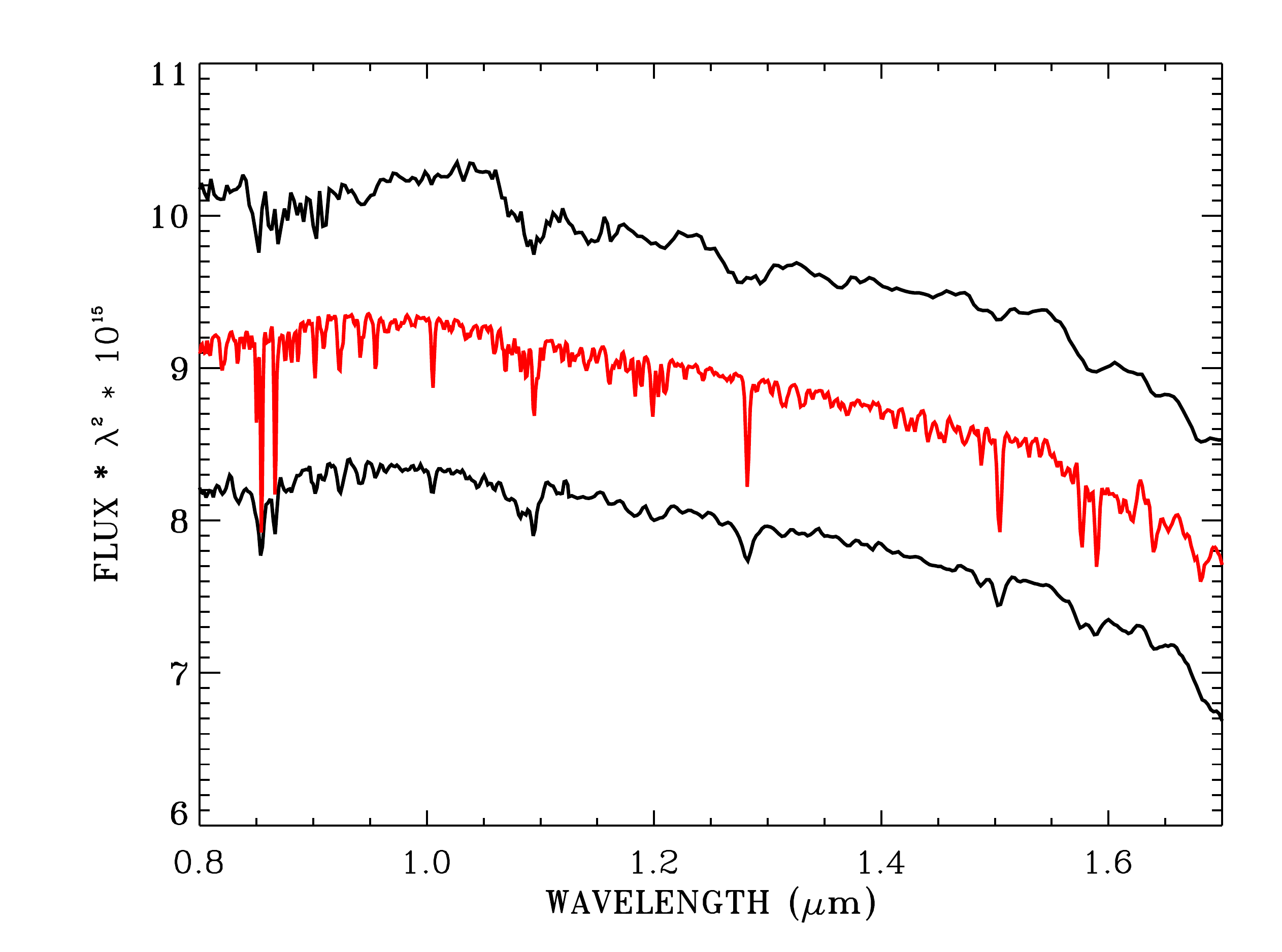}
\caption{Comparison of the WFC3 IR SED to NICMOS for P330E. The BOSZ model for
P330E is shown in red, while the WFC3 (below) and NICMOS (above) fluxes are
multiplied by 0.9 and 1.1, respectively. Notice the good correspondence between
the spectral features for the model and WFC3 SED, while the correspondence is
poor for the NICMOS data. \textbf{The units of the Y-axis are the flux in 
10$^{-15}$ erg
s$^{-1}$ cm$^{-2}$ \AA$^{-1}$ multiplied by the wavelength squared in \micron\
to flatten the traces.}} \label{p330e} \end{figure}

Differences between these new WFC3 IR results and NICMOS are the higher
precision in the wavelength calibration (see Figure~\ref{spex}) and improved
resolution for WFC3 IR compared to NICMOS. The biggest differences between the
new WFC3 results and the NICMOS fluxes is for 2M055914, where Figure~\ref{spex}
illustrates some typical diferences. Figure~\ref{p330e} compares the WFC3 and
NICMOS SEDs to the model for P330E and shows the improved resolution of the WFC3
compared to NICMOS. Stellar absorption lines that match the model are evident in
the WFC3 SED but not in the lower resolution NICMOS data. Visibility of stellar
features enables essential adjustments to the wavelength vector to avoid
wavelength errors that translate directly to flux errors.

\begin{figure}[htb!]			%fig8
\centering 
\includegraphics[width=5.5in,trim=0 50 0 0]{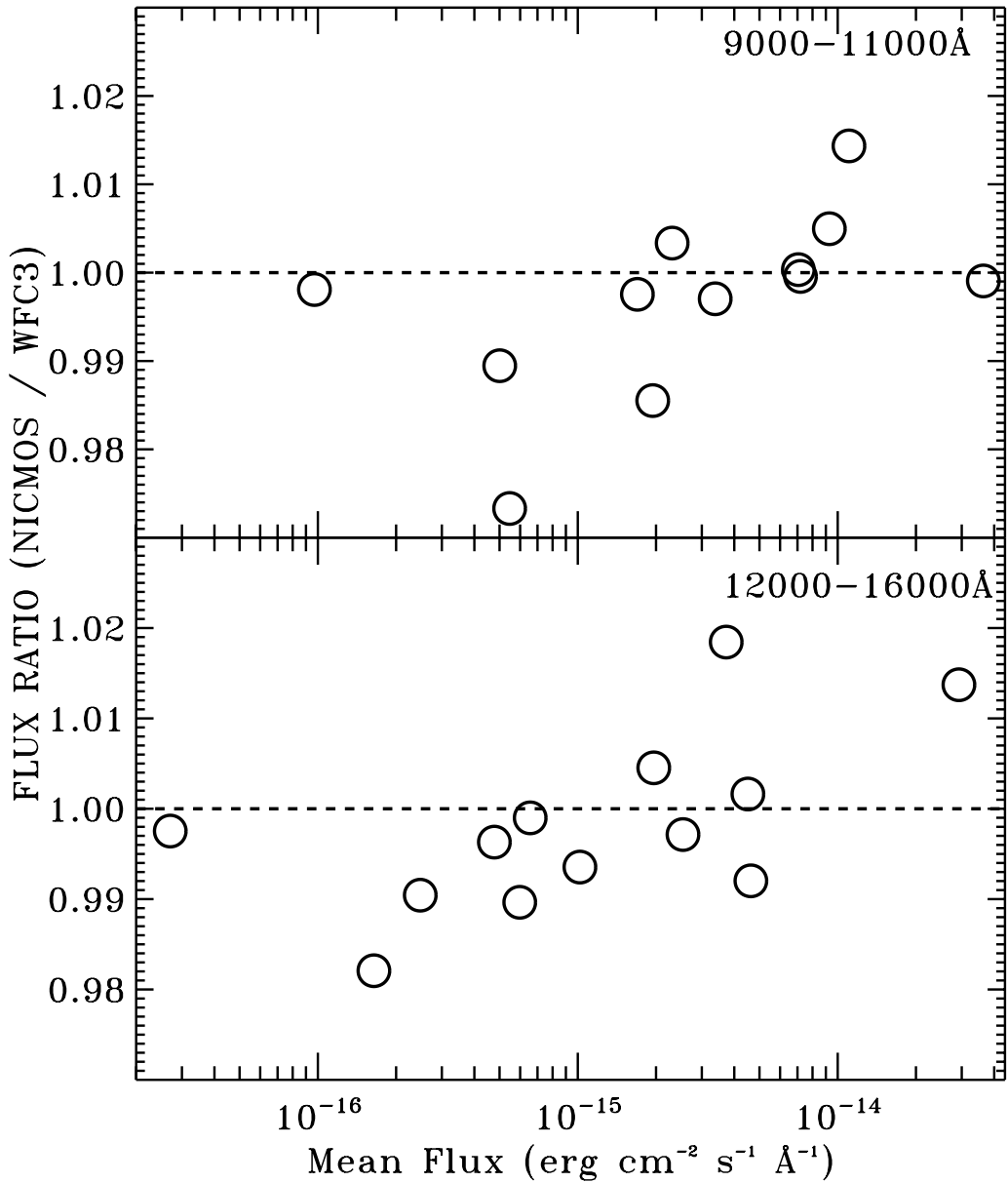} 
\caption{\textbf{Ratio of NICMOS to WFC3 fluxes for G102 (upper panel, 12 stars) and G141 (lower panel, 13 stars). The flux for each star is averaged over the broad bands of 9000-11000~\AA\ for G102 and over 12000-14000~\AA\ for G141.
The X-axis is the same average WFC3 flux used for the denominator of the 
ratios.}}
\label{wfcnic}
\end{figure}

Both the WFC3 and NICMOS flux distibution are compared to STIS in the overlap
region at 8300--9700~\AA\, where the \textit{rms} deviations of the flux ratios
from unity are 0.8\% and 0.6\% for NICMOS/STIS and WFC3/STIS, respectively. At
the longer wavelengths, Figure~\ref{wfcnic} shows the ratios of NICMOS/WFC3
fluxes for the WFC3 G102 and G141 regions at 9000--11000 and 12000--16000~\AA,
respectively. The trends in Figure~\ref{wfcnic} suggest that the CRNL
correction for NICMOS is underestimated by \citet{bohlin06}, but only by
$\sim0.8\%$ per dex in both wavelength regions. These errors of
$\sim0.8\%$ per dex are within the uncertainties of the \citet{bohlin06} NICMOS
non-linearity corrections of $4.8\%$ per dex at 10000~\AA\ and $2.6\%$
per dex at 14000~\AA.

\subsection{Model Fits to SEDs with WFC3 Fluxes}		%6.2

Improvements to the model fits accrue from the improved WFC3 SEDs. 
\citet{bohlin2017} modeled the nine K--A stars from Table~\ref{table:obs} in
order to extrapolate the observed HST SEDs to 32~\micron\ for JWST calibration.
The new SEDs that include the WFC3 IR grism data are in CALSPEC and have been
re-fit with the BOSZ model grid using the same chi-square ($\chi^2$) technique
as \citet{bohlin2017}. The search for a minimum $\chi^2$ proceeds over the
four parameters $T_\mathrm{eff}$, $\log g$, $\log z$, i.e. [M/H], and
interstellar reddening from the dust E(B-V). To calculate the best fit, the data
are binned to avoid strong lines and regions of large line-blanketing, where the
model calculations are less precise. The bins used by \citet{bohlin2017} also
avoided the NICMOS region below 1.3~\micron, where the non-linearity correction
is the largest. However, the non-linearity correction for the WFC3 IR grisms is
wavelength independent. The WFC3 wavelength bins used for cool and for hot stars
appear in Table~\ref{table:bins}, and the new parameters of the model fits for
nine stars appear in Table~\ref{table:fits}.

\begin{figure}[htb!]			%fig9 chifit/modlcont.pro
\centering 
\includegraphics[width=6.5in]{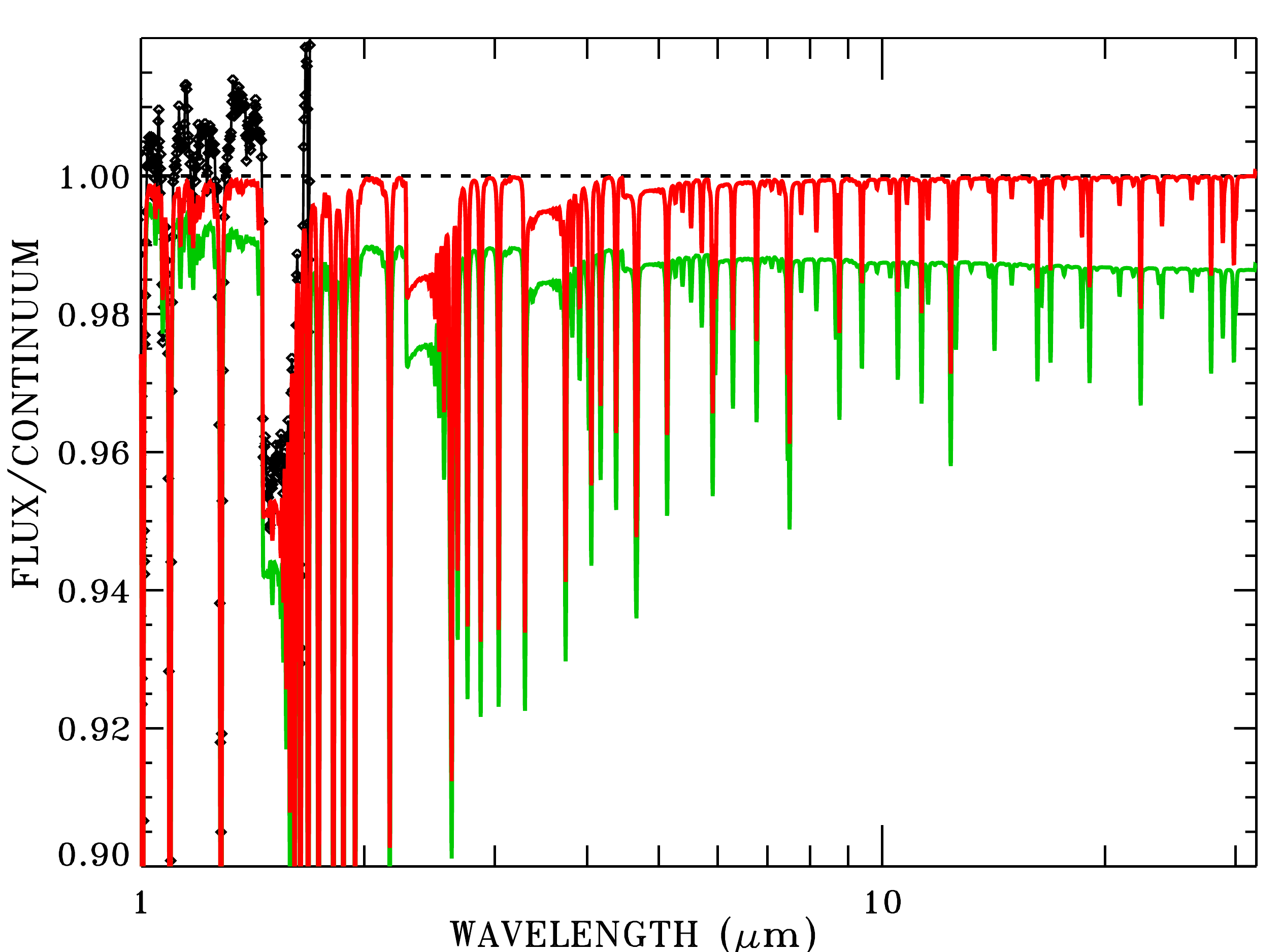} 
\caption{Black data points: WFC3 IR grism flux at 1--1.7~\micron\ for 1808347.
Red: BOSZ model fit to the STIS+WFC3 SED. Green: BOSZ model fit to just the STIS
SED. Fluxes for all three spectral traces are divided by the same model
continuum for the red fit to the STIS+WFC3 SED in order to illustrate
differences at the 1\% level. The WFC3 SED and its red fit agree within 1\% over
most of the observed range. At the longer wavelengths, the green fit to STIS
alone is more than 1\% lower than the improved red fit to the full SED that
includes the new WFC3 IR grism results.} \label{1808} \end{figure}

The biggest differences among the nine modeled IR SEDs are for C26202 and
1808347 with increases of $\sim$1\% longward of 2~\micron. An example of the
improvement over the model fit to only the STIS data appears in
Figure~\ref{1808} for 1808347. The BOSZ model parameters for the fit to the
WFC3+STIS observations are in Table~\ref{table:fits}, while the STIS only fit
parameters are $T_\mathrm{eff}$=7910~K, $\log g$=3.85, $[M/H]$=-0.61, and
E(B-V)=0.024. The red model fit is within 1\% of the new WFC3 IR SED over most
of its 1--1,7~\micron\ range. The old (green) model fit is more than 1\% low
with respect to the WFC3 data and with respect to the new model fit at at the longer wavelengths.

%\newpage
\begin{deluxetable}{cccc}     %Table2 sens.table
\tablewidth{0pt}
\tablecolumns{4}
\tablecaption{\label{table:sens} \mbox{WFC3 IR Grism Flux Calibration~\textit{S}}}
\tablehead{
\multicolumn{2}{c}{~~~~~G102} &\multicolumn{2}{c}{~~~~~~~~G141}\\
\colhead{Wavelength (\AA)} &\colhead{\textit{S}\tablenotemark{a}}
	&\colhead{~~~Wavelength (\AA)} &\colhead{\textit{S}\tablenotemark{a}}}
\startdata
  7608.0000  &1.288E+15    &10000 	 &8.590E+14  \\
  7624.0000  &1.679E+15	   &10025 	 &9.605E+14  \\
  7640.0000  &2.107E+15	   &10050 	 &1.047E+15  \\
  7656.0000  &2.645E+15	   &10075 	 &1.125E+15  \\
  7672.0000  &3.422E+15	   &10100 	 &1.191E+15  \\
  7688.0000  &4.465E+15	   &10125 	 &1.302E+15  \\
\enddata
\tablenotetext{a}{electrons~s$^{-1}$ [erg~s$^{-1}$ cm$^{-2}$
\AA$^{-1}$]$^{-1}$}
\tablecomments{Table~\ref{table:sens} is published in its entirety in a
machine-readable format.}
\end{deluxetable}

\begin{deluxetable}{cccccccc}     %Table3 mrgall.table
%\rotate
\tablewidth{0pt}
\tablecolumns{8}
\tablecaption{\label{table:wfcstub} WFC3 IR SEDs for 19 Stars}
\tablehead{
\colhead{Wavelength (\AA)} &\colhead{Net (electron~s$^{-1}$)}
	&\colhead{Flux\tablenotemark{a}} &\colhead{Stat-err\tablenotemark{a}}
	&\colhead{Sys-err\tablenotemark{a}} &\colhead{No. Obs}
	&\colhead{Exp (s)} }
\startdata
1757132 \\
  7608.0  &4.2500E+01  &3.2997E-14  &7.7682E-16  &3.2997E-16   &8     &211.1 \\
  7624.0  &5.5715E+01  &3.3164E-14  &6.2156E-16  &3.3164E-16   &8     &211.1 \\
  7640.0  &6.7358E+01  &3.1954E-14  &5.1542E-16  &3.1954E-16   &8     &211.1 \\
  7656.0  &8.3295E+01  &3.1480E-14  &4.2731E-16  &3.1480E-16   &8     &211.1 \\
  7672.0  &1.0848E+02  &3.1683E-14  &3.4880E-16  &3.1683E-16   &8     &211.1 \\
  7688.0  &1.4212E+02  &3.1817E-14  &2.8554E-16  &3.1817E-16   &8     &211.1 \\
\enddata
\tablenotetext{a}{erg s$^{-1}$ cm$^{-2}$ \AA$^{-1}$}
\tablecomments{Table~\ref{table:wfcstub} is published in its entirety in a
machine-readable format which includes the SEDs for all 19 stars of
Table~\ref{table:obs}. A portion for the star 1757132 is shown here for guidance
regarding its form and content. Columns Stat-err and Sys-err are the statistical
and 1\% systematic uncertainty estimates.} \end{deluxetable}

%\newpage
\section{Conclusions} %S7

The newly derived WFC3 IR grism SEDs have uncertainties of $\sim$1\% based on
their agreement with the STIS flux distributions. These improved WFC3 IR flux
distibutions have more accurate models and increase confidence that the models
represent the true stellar SEDs to 32~\micron\ with an \textit{rms} precision
approaching 1--2\%. All of the CALSPEC SEDs should be upgraded to include WFC3
IR grism fluxes, especially those stars that are currently lacking NICMOS
coverage or have noisy NICMOS SEDs. \textbf{The more accurate SEDs of the
CALSPEC stars help to better constrain the parameters of the dark energy, where
an absolute color calibration of better than 1\% is required to improve current
results.}

\section*{Acknowledgements}

Scott Fleming entered our HLSP into MAST. Support for this work was provided by
NASA through the Space Telescope Science Institute, which is operated by AURA,
Inc., under NASA contract NAS5-26555. This research made use of the SIMBAD
database, operated at CDS, Strasbourg, France.

~\\ \\  \\  \\
\textbf{ORCID iDs}

Ralph C. Bohlin https://orcid.org/0000-0001-9806-0551 and
Susana E. Deustua https://orcid.org/0000-0003-2823-360X

\newpage
\bibliographystyle{apj}

\bibliography{../../pub/paper-bibliog}

%\newpage
\begin{deluxetable}{cc} 		%table4
\tablewidth{0pt}
\tablecaption{\label{table:bins} Broad Bands in \AA\ for Fitting Stellar Models
	to STIS and WFC3 Grism-IR SEDs}
\tablehead{
\colhead{K--G--F Stars} & \colhead{A-B-O Stars}}

\startdata
              & 1280--1510 \\
              &  1725--2020 \\
              &  2110--2280 \\
              &  2520--2780 \\
3000--3850    &  3000--3200 \\
4000--4260    &  3200--3400 \\
4380--4800     & 3400--3640 \\
4950--5500     & 3750--4400 \\
5500--6000    &  4400--4800 \\
6000--6500     & 4950--5500 \\
6620--7400     & 5500--6000 \\
7400--8400     & 6000--6500 \\
8800--9400     & 6620--7400 \\
9400--10000    & 7400--7900 \\
10200--10750   & 7900--8200  \\
11100--12600   &  9182--9282  \\
13000--14150   &  9290--9480  \\
14150--15300   &  9499--9599  \\
	      &  9600--10000 \\
	      &  10200--10750 \\
	      &  11100--12600 \\
	      &  13000--14150 \\
	      &  14150--15300 \\
\enddata
\tablecomments{These wavelength bands are used for averaging the observed and model fluxes into bins for fitting models from the BOSZ grid according to the method of \citet{bohlin2017} that minimizes $\chi^2$ over four parameters.}\end{deluxetable}

\begin{deluxetable}{lccccc}		%Table5
%\rotate
%\tabletypesize{\scriptsize}
\tablewidth{0pt}
\tablecaption{\label{table:fits} Parameters of the Model Fits}
\tablehead{
\colhead{Star} &\colhead{$T_\mathrm{eff}$} &\colhead{$\log g$}
&\colhead{$[M/H]$} &\colhead{E(B-V)} &$\chi^2$   \\
}
\startdata
1757132            & 7540 & 3.65 & 0.12 & 0.023 &1.14  \\
1802271            & 9060 & 4.00 &-0.47 & 0.019 &0.99  \\
1808347            & 7860 & 3.80 &-0.74 & 0.022 &3.06  \\
BD+60$^{\circ}$1753& 9330 & 3.90 &-0.13 & 0.009 &0.93  \\
C26202             & 6270 & 4.40 &-0.46 & 0.067 &0.23  \\
HD37725            & 8420 & 4.30 &-0.08 & 0.049 &1.37  \\
KF06T2             & 4500 & 1.55 &-0.26 & 0.050 &0.32  \\
P330E              & 5810 & 4.95 &-0.22 & 0.025 &0.40  \\
SNAP2              & 5700 & 4.25 &-0.21 & 0.024 &0.08  \\
\enddata
\tablecomments{Results from fitting BOSZ model atmospheres to the observed stellar SEDs using the method of \citet{bohlin2017} and the wavelength bins of Table~\ref{table:bins}. The parameters of the fit for each star are the effective temperature $T_\mathrm{eff}$, the surface gravity $\log g$, the metallicity $[M/H]$, the interstellar reddening E(B-V), and the chi-square quality of the fit $\chi^2$.} \end{deluxetable}

\end{document}